\documentclass[aps,prl,twocolumn,floatfix]{revtex4}
\usepackage{graphicx}
\usepackage{color}
\usepackage{amsmath}

\newcommand{\be}{\begin{equation}}
\newcommand{\ee}{\end{equation}}
\newcommand{\bea}{\begin{eqnarray}}
\newcommand{\eea}{\end{eqnarray}}

\newcommand{\pd}[2]{\frac{\partial #1}{\partial #2}}

\newcommand{\tr}{{\tilde r}}
\newcommand{\eps}{\epsilon}

\newcommand{\bnh}{{\bf \hat{n}}}

\newcommand{\bF}{{\bf F}}
\newcommand{\bU}{{\bf U}}
\newcommand{\bbe}{{\bf e}}
\newcommand{\bu}{{\bf u}}
\newcommand{\br}{{\bf r}}

\newcommand{\dg}{\dot{\gamma}}
\newcommand{\tdg}{\tilde {\dot{\gamma}}}
\newcommand{\tes}{\tilde \eta_1}
\newcommand{\deta}{\tilde \eta^*_1}
\newcommand{\ter}{\tilde r}
\newcommand{\teta}{\tilde \eta}

\newcommand{\oo}[1]{{\cal O}\!\left( #1 \right)}

\newcommand{\bsig}{\te{\sigma}}
\newcommand{\bdel}{\te{\delta}}

\newcommand{\te}[1]{\mbox{\boldmath$ #1 $}} 
\def\bu{\te{u}}\def\bU{\te{U}}\def\br{\te{r}}\def\bF{\te{F}}\def\b1{\te{1}}

\begin{document}

\title{Active microrheology in the continuum limit:  can the macrorheology be recovered?}
\author{Todd M. Squires}
\affiliation{Departments of Physics and Applied Mathematics, Caltech, Pasadena, CA 91125}
\date{\today}

\begin{abstract}
Active microrheology differs from its passive counterpart in that the probe is actively forced through the material, rather than allowed to diffuse.  Unlike in passive microrheology, active forcing allows the material to be driven out of equilibrium, and its nonlinear response to be probed.  However, this also renders inoperable the fluctuation-dissipation theorem used to justify passive microrheology.  Here we explore a question at the heart of active microrheology:  are its results consistent with macrorheology?  We study a simple model material -- a generalized Newtonian fluid, with a small but arbitrary shear-rate-dependent component -- and derive a general expression for dissipation due to probe motion, which remarkably does not require the non-Newtonian flow to be solved.  We demonstrate that the straightforward application of active microrheology gives results that are {\sl inconsistent} with macrorheology, even when the probe is large enough for material to behave as a continuum, unless the forcing is gentle enough to probe only the linear response.  Regardless, each technique encodes information about the material; if suitably interpreted, the (macro-) constitutive relation can indeed be recovered from the microrheological data.  We emphasize that more, rather than less, information would be obtained if the two methods disagree.
\end{abstract}

\maketitle

For decades, conventional (macroscopic) rheology has been employed to characterize and understand complex fluids and materials \cite{larson99}.  Some materials, however, would be impossible to study using macro-rheometry, due to the difficulty or expense in obtaining sufficient quantities for analysis.  For such materials, techniques in microrheology are being developed \cite{mackintosh99,gisler98,furst03}.  Most microrheology is passive, and relates the thermal fluctuations of translating \cite{mason95,gittes97,crocker00,levine00} or rotating \cite{cheng03,schmiedeberg05} colloidal probes to linear-response, near-equilibrium viscous and elastic moduli through the fluctuation-dissipation theorem. 

More recently, techniques in active and nonlinear microrheology have begun to be explored.  A colloidal probe is optically or magnetically pulled  \cite{lugowski02,habdas04,meyer05,squires05b,khair05} or rotated \cite{bishop04} through the material, and a `microviscosity' 
\be
\eta^*= \frac{F^*}{6 \pi a U^*} \,\,{{\rm or}}\,\, \eta^* = \frac{\tau^*}{8\pi a^3 \Omega^*}
\label{eq:naiveeq}
\ee
is inferred, analogous to the generalized Stokes-Einstein relation assumed in passive microrheology.  (Starred variables will here refer to microrheological quanitites.)  

Active microrheology differs from its passive counterpart in it allows the material microstructure to be driven significantly out of equilibrium, providing a window into the material's nonlinear response.  Such information is not available to passive microrheology, which by nature probes the near-equilibrium properties of the material.  For example, active microrheology is expected to reveal shear/force-thinning and thickening in colloidal suspensions, as is seen macrorheologically \cite{squires05b,khair05}.  

A central question for microrheology is whether or not its results are consistent with those obtained using conventional macro-rheology.  Passive microrheology seems to be on a firmer theoretical footing, as the fluctuation-dissipation theorem can be invoked \cite{levine00}.  Unfortunately, no corresponding theorem exists to ensure agreement between macrorheology and active microrheology.  While such agreement would certainly be reassuring, it is not, strictly speaking, necessary.  In fact, we argue the opposite to be true.  Any disagreement between the two techniques would yield more, rather than less, information about the material:  each probes the material in a different way, and thus encodes different information.  While further work would then be required to properly interpret microrheological data, any discrepancies encode additional physics which could, in principle, be extracted.

Differences between active microrheology and its macro-rheological analog have already been noted.  Studies have thus far employed colloidal suspensions, as they are perhaps the cleanest and simplest systems for theoretical and experimental study.  For example, we demonstrated semi-quantitative agreement between micro- and macro-rheological measurements of the Brownian (microstructural) component of the viscosity of a suspension \cite{squires05b,khair05}.  This agreement carried, however, an important caveat: the data must first be scaled (with colloidal volume fraction $\phi$ for micro vs. with $\phi^2$ for macro) for a meaningful comparison to be made.  Such agreement could only be obtained, however, because the proper scaling and physics for colloidal suspensions was known; for more complex (or unknown) materials, one would not know how to proceed {\sl a priori}.

Another significant difference involves non-continuum effects, which arise when the size of the probe is comparable to microstructural features of the material.  This has been noted in falling-ball rheometry \cite{almog97} and in active microrheology \cite{squires05b}, and is not terribly suprising, since the material is not being probed in the continuum limit.  Two-point microrheology was developed in order to address this issue in the passive microrheological context \cite{crocker00,levine00}; a similar strategy may be useful as well for active microrheology.  More counter-intuitive, perhaps, is that the inferred viscosity depends on whether the probe is pulled at constant force or constant velocity \cite{almog97}, for which a simple physical argument has been offered \cite{squires05b}. 

Here we approach the macro/micro question from the opposite standpoint:  even when the probe is so large that the material behaves as a continuum, does microrheology reproduce macrorheology?  One might assume that it would, since one is probing the continuum properties of the material.  On the other hand, the flow and stress fields established in active microrheology are not viscometric, but decay with distance from the probe.  By contrast, (macro-) rheologists have long emphasized and utilized viscometric flows \cite{bird87a}.  In light of this distinction, any macro/micro agreement should appear surprising.

To address this question in a simple, clear, and general fashion, we consider a probe that is driven through a generalized Newtonian fluid, whose viscosity
\be
\eta = \eta_0+ \epsilon \eta_1(\dot{\gamma}),
\label{eq:viscosity}
\ee
is nearly Newtonian, but which contains a small, shear-dependent component $\eta_1$, whose functional form is arbitrary (although by construction $\eta_1(0) =0$). 

In what follows, we derive a general relation for the dissipation due to a probe moving steadily in such a material.  This allows the microviscosity $\eta_1^*$ to be obtained {\sl without} solving for the non-Newtonian component of the flow field.  We demonstrate that the micro- and macro- viscosities are in fact different, so that a naive application of active microrheology via (\ref{eq:naiveeq}) would be inappropriate unless the forcing was extremely weak.  Instead, we show that $\eta_1^*$ is related to $\eta_1$ through a weighted average that depends on the shape and properties of the probe.  This relation can be inverted numerically in general, and exactly in at least one case.

We pose an expansion for the fluid velocity $\bu = \bu_0+\eps \bu_1 + ...$, where the leading-order solution simply solves Stokes' equations
\bea
\nabla\cdot \bsig_0=\eta_0 \nabla^2 \bu_0 - \nabla p = 0,\,\,\nabla\cdot\bu_0 = 0\\
\bu_0(\Gamma)=\bU,\,\,\bu_0(\br\rightarrow\infty)\rightarrow {\bf 0}
\eea
subject to no-slip boundary conditions.  The $\oo{\epsilon}$ non-Newtonian correction then obeys
\bea
\nabla\cdot \bsig_1=\eta_0 \nabla^2 \bu_1 - \nabla p_1=-{\bf f}_1,\,\,\nabla\cdot\bu_1 = 0\\
\bu_1(\Gamma)={\bf 0},\,\,\bu_1(\br\rightarrow\infty)\rightarrow {\bf 0}.
\eea
Here $ \bsig_n = -p_n \bdel + \eta_0 \bbe_n$ is the stress tensor associated with the  $n^{{\rm th}}$-order velocity field, $\bbe_n =  \nabla \bu_n +(\nabla \bu_n)^T$ is the  $n^{{\rm th}}$-order rate of strain tensor, and ${\bf f}_1$ represents the body force $
{\bf f}_1 = \nabla\cdot(\eta_1\bbe_0)$.

One can show that the dissipation due to a body moving through a generalized Newtonian fluid is given by
\be
\bF^*\cdot\bU^* = \frac{1}{2}\int \eta(\dot{\gamma}) \bbe:\bbe dV\equiv \frac{1}{2}\int \eta(\dot{\gamma})  \dot{\gamma}^2 dV.
\ee
Using (\ref{eq:viscosity}), the dissipation is given to $\oo{\eps}$ by
\be
\bF^*\cdot\bU^* = \frac{1}{2}\int (\eta_0+\eps \eta_1)\bbe_0:\bbe_0 dV + \eps\int \eta_0 \bbe_1:\bbe_0 dV.
\ee 
We now show the second integral to vanish.  Writing $\eta_0 \bbe_1 = \bsig_1 + p_1\bdel$, we note that the isotropic ($p_1$) term vanishes upon contraction with (traceless) $\bbe_0$, giving $\int \eta_0 \bbe_1:\bbe_0 dV =  \int \bsig_1:\bbe_0 dV$.  Expanding $\bbe_0$ and using the divergence theorem gives
\be
\int  \eta_0 \bbe_1: \bbe_0 dV=2\int \bnh\cdot   \bsig_1 \cdot \bu_0 dA +2 \int \bu_0 \cdot {\bf f}_1 dV,
\label{eq:vanishes}
\ee
where the surface integral is taken over the body and at infinity.  
Finally, a generalized reciprocal theorem for Stokes flow that accounts for body forces $\nabla \cdot \bsig = -{\bf f}$,
\be
\int \bnh\cdot  \bsig_1 \cdot \bu_0 dA + \int \bu_0 \cdot {\bf f}_1 dV=\int \bnh\cdot   \bsig_0 \cdot \bu_1 dA + \int \bu_1 \cdot {\bf f}_0 dV,\ee
shows (\ref{eq:vanishes}) to vanish identically, since ${\bf f}_0 = 0$ everywhere and $\bu_1={\bf 0}$ on all surfaces.
We thus arrive at
\be
\bF^*\cdot\bU^* = \frac{1}{2}\int (\eta_0+\eps \eta_1)\bbe_0:\bbe_0 dV + \oo{\eps^2}.
\ee 
An analogous relation holds for rotating probes.  Significantly, {\sl only} the Stokes flow around the probe is required to obtain the dissipation to $\oo{\eps}$.

A straightforward application of (\ref{eq:naiveeq}) would yield 
\be
\eps \eta^*_1 = \frac{\eta^*-\eta_0}{\eta_0} = \eps \frac{\int \eta_1(\dot{\gamma}_0) \dot{\gamma}^2_0 dV}{\eta_0\int \dot{\gamma}^2_0 dV}.
\label{eq:effvisc}
\ee
Even in this simple example, the micro- and macro-viscosities are not equal unless $\eta_1$ is nearly constant over the integration volume -- {\sl i.e.} $\dg$ is very small (linear response, as with Brownian probes) or asymptotically large.  Rather, $\eta_1^*$ represents a weighted average of $\eta_1$, reflecting the non-viscometric flow around the probe.

From now on, we scale lengths by the probe size $a$, $\eta_1$ by $\eta_0$, $\dg_0$ by the maximum (Stokes) shear rate $\dg_{{\rm max}}$, and denote dimensionless variables with tildes.  We consider $\dg$ in $\eta_1$ to be parametrized by $\dg_c$, and denote $\delta = \dg_{{\rm max}}/\dg_c$.  With this notation, Eq. \ref{eq:effvisc} becomes
\be
\deta(\delta)= \frac{\int \tes(\delta \tdg_0) \tdg^2_0 d \tilde V}{\int \tdg^2_0 d \tilde V}.
\label{eq:nondvisc}
\ee
We interpret the microviscosity in terms of $\delta$, the (scaled) maximum shear rate from the Stokes flow solution.

An obvious question is whether one can recover $\tes$ uniquely from $\deta$.  If not, then different macroviscosities could yield the same micro-viscosity $\deta$.  This would imply in turn that their difference $\teta_D$ would obey
\be
\int \teta_D(\delta \tdg_0) \tdg^2_0 d \tilde V=0.
\ee
Because $\teta_D$ is a function of $\delta$ and $\tdg^2_0$ is positive definite, all coefficients of a power series in $\delta$ of $\teta_D$ must be zero for this relation to hold.  This would imply $\teta_D\equiv 0$, and we conclude that (\ref{eq:nondvisc}) is indeed invertible.

Having arrived at the general relation, we explore illustrative examples of consitutive relations $\tes$ and probe forcing.  The simplest example, a fluid with a slight power-law viscosity, is recovered microrheologically as 
\be
\tes=  \delta^n \tdg^n,\,\,\deta = A_n \delta^n, \,\,{{\rm where}}\,\,
A_n =  \frac{\int \tdg^{2+n}_0 d \tilde V}{\int \tdg^2_0 d \tilde V}.
\label{eq:an}
\ee
The functional form for $\deta(\delta)$ is identical to $\tes(\dg)$, but with a probe-dependent prefactor $A_n$.  For example, a translating sphere (TS) has shear rate
\be
\tdg^2 = \frac{1 + (2-6\ter^2+3\ter^4)\cos^2 \theta}{\ter^8},\,\, \dg_{{\rm max}} = \frac{3}{\sqrt{2}}\frac{U}{a},
\label{eq:sphereshear}
\ee
giving coefficients $A_n$ shown in Table \ref{tab:powerlaw}. Coefficients for translating bubble (TB),  
\be
\tdg = \ter^{-2} \cos \theta, \,\,\dg_{{\rm max}}=\sqrt{6} U/a,
\ee
and rotating sphere (RS) probes
\be
\dot{\gamma} =\ter^{-3}\sin \theta, \,\,\dg_{{\rm max}} = 3\sqrt{2}\Omega.
\label{eq:shearrot}
\ee
are also shown in Table \ref{tab:powerlaw}.

\begin{table}
\begin{center}
\begin{tabular}[t]{|c|c|c|c|}
\hline
$n$
&    Translating&Translating&Rotating\\
&Sphere&Bubble&Sphere\\
\hline
$\dg_{{\rm max}}$ & $3U/\sqrt{2}a$ & $\sqrt{6}U/a$ & $3\sqrt{2}\Omega$\\
\hline
$A_1$ 	& 0.210	& 0.250 	& 0.442	\\
$A_2$ 	& 0.092	& 0.120	& 0.267	\\
$A_3$	& 0.056	& 0.071	& 0.184	\\
$A_4$ 	& 0.037	& 0.048	& 0.137	\\
$A_5$	& 0.027	& 0.034	& 0.107	\\
\hline
\end{tabular}
\caption{Prefactors (from eq. \ref{eq:an}) for the power-law viscosity $\deta$ obtained active microrheology, given a macro-viscosity $\tes\sim\dg^n$.  Three different methods of forcing are presented:  a translating sphere and bubble, and a rotating sphere.}
\label{tab:powerlaw}
\end{center}
\end{table}

Even for the simplest model fluid, $A_n\neq 1$, and so the microviscosity differs from the macroviscosity.  Furthermore, general viscosities $\tes$ can be expanded as power series in the low-$\delta$ limit, 
\be
\tes \sim \sum_{n=1}^\infty B_n \delta^n \tdg^n,
\ee
whereas an expansion of the microviscosity would give
\be
\deta \sim \sum_{n=1}^\infty B_n A_n \delta^n.
\ee
Although $\deta \neq \tes$, the macro-viscosity coefficients $B_n$ could be determined microrheologically, since $A_n$ can be considered known.  Again, the straightforward application of active microrheology is {\sl inconsistent} with macrorheology, except in the linear-response limit ($\dg\rightarrow 0$).  Nonetheless, the connection between the two exists and can be extracted with proper interpretation.

\begin{figure}
\begin{center}
\centerline{
\includegraphics[height=1.6in]{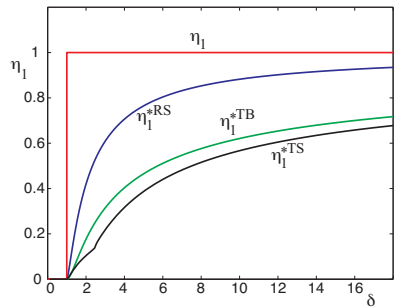}
}
\caption{
\label{fig:stepvisc}
Microviscosities obtained from a fluid with abruptly shear-thickening viscosity $\eta_1$ using a rotating sphere (RS), translating sphere (TS) and translating bubble (TB).  Although the shear-thickening is seen to begin at the correct shear rate $\delta$, the microviscosities approach the macro-viscosity only slowly.}
\end{center}
\end{figure}

We now turn to the opposite limit, when the probe is pulled strongly enough to sample interesting features in $\tes$.  The simplest illustrative function for $\tes$ is a fluid that thickens abruptly, via a step function,
\be
\tes = \Theta(\tilde \dg - 1/\delta).
\ee
For $\delta < 1$, nowhere does the material thicken, and one recovers $\deta(\delta<1)=0$.    When $\delta>1$, some volume of the material thickens, as seen in Fig. \ref{fig:stepvisc}.  Eq. (\ref{eq:nondvisc}) can be evaluated exactly for TB and RS forcing,
\bea
\eta_1^{*TB} &=& 1 -\frac{6}{5 \delta^{1/2}} + \frac{1}{5\delta^3},
\label{eq:steptb}\\
\eta_1^{*RS} &=& \frac{(4\delta^2-1)\sqrt{\delta^2-1} - 3\delta^2\sin^{-1}\left(\sqrt{\delta^2-1}/\delta\right)}{4\delta^3}.
\label{eq:steprs}
\eea
Although all three probes do show the onset of shear thickening at the correct $\delta$, they approach the `true' value only slowly:  like $\delta^{-1/2}$ for translating bodies and $\delta^{-1}$ for rotating. The reason for this slow decay is purely geometric:  the fluid thickens until $\tdg\sim 1/\delta$, which occurs at $r\sim\delta^{1/2}$ ($r\sim\delta^{1/3}$) for translating (rotating) probes.  The integrand scales like $\delta^{-2}$, and the volume element scales like $ \delta^{3/2}$ ($\delta$) for translations (rotations), leaving a $\delta^{-1/2}$ ($\delta^{-1}$) decay. Unfortunately, such slow decays are ubiquitous in active microrheology, and are remnants of interesting features at intermediate values of $\delta$.

Rotating probes are marginally better at reproducing the macro-viscosity because the faster spatial decay of $\dg$ effectively weights the average closer to the probe.  Taking this strategy to its logical extreme (a shear rate that decays infinitely quickly) would yield a microviscosity that averages strictly over the probe surface, and does not suffer from these long-range geometric effects.  

In fact, we demonstrate that this idea can actually be realized.  Two of the probe motions described above (TB and RS) are special, in that the flow for different-sized probes are identical if their forces/torques are the same (meaning $aU_a=bU_b$ and $a^3 \Omega_a = b^3 \Omega_b$).  Taking the difference between dissipation integrals in (\ref{eq:nondvisc}) at the appropriate $\delta_a$ and $\delta_b$ yields a volume integral over the {\sl difference} in probe volumes, since the integrands are identical outside of the larger probe.  For example, using this technique for the TB mode gives
\be
\frac{(F^*)^2}{4\pi\eta_0} \left(\frac {\tilde \eta_1^{*TB} (\delta_a)}{a}-\frac{\tilde \eta_1^{*TB}(\delta_b)}{b}\right)=\frac{1}{2}\int_a^b \eta_1(\dg^2)\dg^2 dV,
\ee
where $\delta_a = (b^2/a^2)\delta_b$ to enforce $F^*_a=F^*_b\equiv F^*$.
The limit $a\rightarrow b$ yields an expression for the surface-averaged macroviscosity,
\be
2\delta \pd{\tilde \eta_1^{*TB}}{\delta} + \tilde \eta_1^{*TB} = \frac{3}{4\pi}\int_{\tr=1} \tes(\delta \tdg)\tdg^2 d\Omega,
\ee
where $\delta_b = \delta$.  Since $\delta \tdg =\delta \cos\theta \equiv u$ on $\ter=1$, we perform the integral over $\phi$ to obtain
\be
2\delta \pd{\tilde \eta_1^{*TB}}{\delta} + \tilde \eta_1^{*TB} = \frac{3}{\delta^3} \int_0^\delta \tes(u)u^2 du,
\ee
which can be inverted exactly to reveal a remarkable relation between the micro- and macro viscosities,
\be
\tes(\delta) = \frac{1}{\delta^2} \pd{}{\delta}\left[\frac{\delta^3}{3}\left(2\delta \pd{\tilde \eta_1^{*TB}}{\delta} + \tilde \eta_1^{*TB}\right)\right].
\label{eq:holycrap}
\ee
One can verify (\ref{eq:holycrap}) for the step-function and power-law viscosities considered above.
The analogous surface-averaged viscosity relation for RS forcing,
\be
\pd{(\tilde \eta_1^{*RS} \delta)}{\delta} = \frac{3}{2}\int_{0}^\pi \tes(\delta \sin\theta)\sin^3\theta d\theta,
\ee
does not appear to have such a simple inversion.

Several important points emerge from this simple analysis.  Most significantly, the straightforward application of active microrheology will not yield results consistent with macrorheology, unless the applied force/torque are small enough that the material microstructure is not significantly deformed.  In the nonlinear regime, the micro- and macro viscosities for weakly shear-dependent fluids are related through weighted averages that can be inverted uniquely.  This result holds only for weakly non-Newtonian fluids, whose velocity field is approximately Stokes flow, and whose non-Newtonian component is a small perturbation.  Whether similar results exist for general materials remains an open question.  

We close with a brief reflection on the agreement between (active) microrheology and macrorheology, by posing a broader question about constitutive relations (whose validity we have assumed without comment).  Such relations generally assume viscometric flow and large-amplitude, steady deformation \cite{bird87a}.  It is possible that probes in active microrheology do not incur strains that are significant or steady enough for such constitutive relations to hold, despite the magnitude of $\dg$.  While the relations derived above may be  confidence-inspiring, we feel it would be more interesting if they were experimentally shown {\sl not} to hold.  In that case, active microrheology would in fact be providing information that is inaccessible to macrorheological techniques.   Since most flows in the real world are not viscometric, it would be interesting and important to investigate what accuracy (or relevance) constitutive relations hold in more general situations.    Active microrheology seems ideally suited for such investigations.



\end{document}